\documentclass[12pt]{iopart}

\usepackage{graphicx}
\usepackage{iopams} 
\usepackage{setstack}

\begin{document}

\title{An invisible quantum tripwire}
\author{Petr M Anisimov\footnote{Author to whom any correspondence should be addressed.}, Daniel J Lum, S Blane McCracken, Hwang Lee and Jonathan P Dowling}
\address{Hearne Institute for Theoretical Physics and Department of Physics and Astronomy \\
Louisiana State University, Baton Rouge, LA 70803 }
\ead{petr@lsu.edu}

\begin{abstract}
We present here a quantum tripwire, which is a quantum optical interrogation 
technique capable of detecting an intrusion with very low probability of the 
tripwire being revealed to the intruder. Our scheme combines interaction-free 
measurement with the quantum Zeno effect in order to interrogate the presence 
of the intruder without interaction. The tripwire exploits a curious nonlinear 
behaviour of the quantum Zeno effect we discovered, which occurs in a lossy 
system.  We also employ a statistical hypothesis testing protocol, allowing us 
to calculate a confidence level of interaction-free measurement after a given 
number of trials. As a result, our quantum intruder alert system is robust 
against photon loss and dephasing under realistic atmospheric conditions and 
its design minimizes the probabilities of false positives and false negatives 
as well as the probability of becoming visible to the intruder. 
\end{abstract}

\submitto{\NJP}

\maketitle

\section{Introduction}
Interaction Free Measurement (IFM) originated with the Elitzur-Vaidman ``Bomb'' 
gedanken experiment that showed it was possible to detected a single-photon, 
hair-triggered bomb in an interferometer --- without setting it off --- by 
exploiting single particle interference combined with the presence of quantum 
``which-path'' information \cite{vaidman}. The original bomb protocol had a 
success probability of only 25\%. (In another 50\% of the runs the bomb was 
detonated, and in the remaining 25\% no information about the bomb was 
obtained.)  The protocol was improved upon by Kwiat, et al., who 
combined lossless IFM with a multi-pass quantum Zeno effect \cite{kwiat}. 
In our work presented here, we discovered a curious nonlinear behaviour of 
photon's transmission in a Zeno enhanced but lossy IFM apparatus. This discovery
leads us to an IFM protocol robust against photon loss and dephasing. In 
addition, we recast the entire protocol in terms of statistical hypothesis 
testing, allowing us to quantify the operation of the device as a reliable yet 
undetectable intruder alert system --- the invisible quantum tripwire.

\section{Interaction free measurement}
The Elitzur-Vaidman ``Bomb'' gedanken experiment posits that there exists a 
bomb with a single-photon sensitive detonator and the goal is to optically 
detect the presence of such a 
bomb without detonation. In contrast to the expectations of
the classical approach, where such a goal could not be reached, quantum optics 
allows for a solution --- measurement without interaction.
\begin{figure}[t]
\begin{center}
\includegraphics{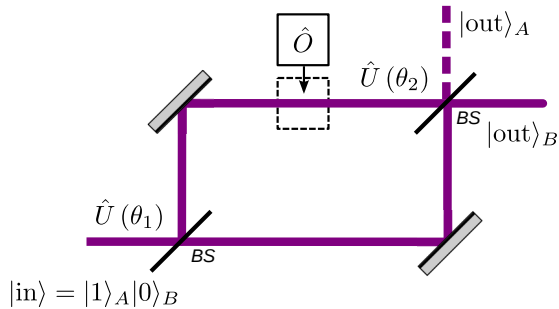}
\end{center}
\caption{\label{fig:SimpleSetup} A lossless Mach-Zehnder interferometer 
in a dark port arrangement, $\theta_1+\theta_2=\pi/2$, and a zero phase
difference between its arms, constitutes a simple IFM setup with
efficiency $\eta\le1/2$. This scheme allows for interaction-free hypotheses 
testing of a path being blocked ($h_1$) or it being clear ($h_0$).}
\end{figure}

This measurement is based on a fascinating property of a single photon to 
interfere with itself while being indivisible.
Imagine a lossless Mach-Zehnder interferometer (MZI) with beam splitters
described by a two mode coupling matrix
 \begin{equation}
 \hat{U}\left(  \theta_{i}\right)  =%
\left[\matrix{%
\cos\theta_{i} & -\sin\theta_{i}\cr
\sin\theta_{i} & \cos\theta_{i}\cr
}\right]  
 \end{equation}
and the possibility of a photon-sensitive object to be placed in the detection
arm (see figure \ref{fig:SimpleSetup}). This detection arm stays invisible
to the object for as long as a photon has not been absorbed by the object.
There are two possible scenarios: the path is blocked or it is clear.
If the path is clear, a single photon, after the first beam splitter $\hat{U}(\theta_1)$, 
can travel both arms of an interferometer and interfere with itself 
at the second beam splitter $\hat{U}(\theta_2)$. Under a proper choice of beam splitters,
$\theta_1+\theta_2=\pi/2$, and a zero phase difference, such an interference will result in 
zero probability of the photon to leave the MZI in mode A (dark port),
that is $P_0(D)=0$. If the path is blocked by an object $\hat{O}$, there is a definite 
destruction of the interference as well as the probability for an object to 
absorb a photon, $P_{1}(A)=\sin^2\theta_1$. Loss of the photon tells us that an object is 
there, but this is a measurement with an interaction. Without interference 
there exists a non-zero probability for a photon exiting the MZI through the 
dark port, $P_1(D)=\cos^2\theta_1\cos^2\theta_2$. Detection of a photon in a dark port constitutes 
a measurement without an interaction. The efficiency of a given measurement is 
$\eta=P_1(D)/[P_1(D)+P_1(A)]$, since an object is detected with probability $P_1(D)+P_1(A)$, 
while detection without interaction is carried out with probability $P_1(D)$.
In the presented setup, there is a limit on the highest efficiency
$\eta=\cos^2\theta_1/(1+\cos^2\theta_1)\le1/2$, which is achieved at the limit where $P_1(D)\rightarrow 0$,
$P_1(B)\rightarrow P_0(B)=1$ and single trial detection becomes improbable.

\section{Invisible hypothesis testing}
Clearly, these two scenarios correspond to two hypotheses $h_1$ and $h_0$
that an IFM apparatus tests for without interaction. These two hypotheses
hold equal statistical weight (symmetric testing) and are described by possible 
outcomes $b=\{A,B,D\}$ and their corresponding probabilities $P_1(b)$ and $P_0(b)$ for the 
first and second hypothesis, respectively. Due to probabilistic nature of the
outcomes, there is always a chance of a false positive.
This error of choice after a single trial is limited by the classical Chernoff 
bound \cite{ISI:A1952UM02500001}:
\begin{equation}
\label{eq:cherbound}
P_{e}\leq \frac{1}{2}\underset{s\in [0,1]}{\min }\sum _{b}P_0^{s}(b)P_1^{1-s}(b).
\end{equation}
Meanwhile, the classical Chernoff bound on the error of choosing the wrong 
hypothesis after $M$ trials scales
exponentially, $P_{e}<\frac{1}{2}e^{-MC\left(P_0,P_1\right)}\equiv P_{e}^{\rm max}$, where $C\left(P_0,P_1\right)=\underset{s\in [0,1]}{-\min }\ln \left(\sum _{b} 
P_0^{s}(b)P_1^{1-s}(b)\right)$ is known as 
the Chernoff distance.

Our IFM apparatus performs interaction free hypotheses testing based on
three possible outcomes: ($b=A$) the probability of absorption because of photon
loss or a measurement with an interaction, ($b=D$) the probability of an IFM, 
and ($b=B$) the probability of learning nothing where the photon exits through 
the bright port of the interferometer. The importance of no photon loss without
an object, $P_{0}(A)=0$, and the dark-port condition, $P_0(D)=0$, becomes now obvious 
in the light of \Eref{eq:cherbound}. These assumptions ensure that the error 
of false acceptance comes from the probability of a photon to exit through the 
bright port in the presence of an object $P_1(B)=\cos^4\theta_1$ and is equal to $P_{e}=\frac{1}{2}P_1(B)$ 
due to a 50-50 chance of wrongly choosing after such an outcome. 

The error of false acceptance in a lossless MZI with a dark port is minimized
by an increase of the first beam splitter's reflectance ($\theta_1\rightarrow \pi/2$). It means that
all the photons are routed into the detection arm. Hence, interaction with an 
object becomes unavoidable and the photon path becomes visible. In the opposite 
case, $\theta_1\rightarrow 0$, the probability of an interaction with the object is significantly 
reduced, at the expense of high statistical error. In order to compensate for 
the increased statistical error, multiple trials are required. For the photon 
path to stay invisible to the object, every photon must be received at the 
output, which happens with the probability $\bar{P}_{\rm{vis}}=\exp(-MC_{\rm{vis}})$, where the visibility
distance, $C_{\rm{vis}}=-\ln\cos^2\theta_1$, is introduced for an easy comparison with the 
Chernoff distance, $C(P_0,P_1)=-2\ln\cos^2\theta_1$. Judging by these distances, it is 
possible for the detection to be hidden from the object, $\bar{P}_{\rm{vis}}\gg 0$, while 
revealing the presence of the object with a high level of certainty $P_e\rightarrow 0$.
Sadly, any deviation from the ideal setup --- such as loss, phase shifts, or 
non-perfect dark port arrangement makes the Chernoff and visibility distances 
comparable; thus effectively preventing the invisibility of 
a tripwire based on IFM in a setup presented in figure \ref{fig:SimpleSetup}.

\begin{figure}[t]
\begin{center}
\includegraphics{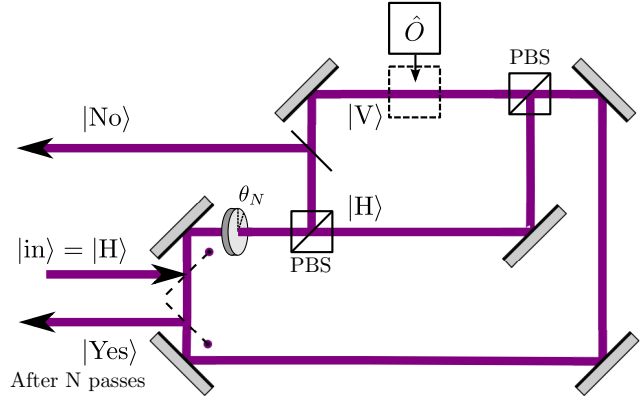}
\end{center}
\caption{\label{fig:ZenoBasedIFM} IQT apparatus based on a $N$-pass IFM in the 
polarization interferometer. With each pass, a photon's polarization is rotated 
by an angle $\theta_{N}$. The presence of an object prevents accumulation of
polarization rotation and is similar to the quantum Zeno effect \cite{kwiat95b,kwiat99}.
An additional beam splitter inside the polarization interferometer models 
unavoidable loss in the arm accessible by the object as well as controlled loss
that is adjusted to provide best performance of the IQT apparatus.}
\end{figure}

\section{Invisible tripwire}
Nevertheless, an invisible quantum tripwire (IQT) is possible. We realize it
through a combination of an efficient IFM apparatus and a proper interrogation 
technique.
A possible IQT apparatus is presented in figure~\ref{fig:ZenoBasedIFM} and
is based on a $N$-pass IFM apparatus, which offers improved efficiency $\eta$ due to 
the quantum Zeno effect \cite{kwiat95b,kwiat99}. 
A crucial part of IQT apparatus is,
however, a quantum interrogation technique that deals
much better with high sensitivity of the $N$-pass IFM to photon loss 
\cite{rudolph}, as well as eliminates the dark-port condition.
 This technique is based on the partial Zeno effect and actually adds a 
controllable amount of loss to the detection arm by means of a beam splitter 
with
tunable reflectivity. Any attempt to register a photon (that constitutes a tripwire)
 as well as crossing 
the path of a photon, would immediately engage the quantum Zeno effect resulting 
in drastic reduction of the photon loss. This effect will increase the rate at 
which photons exit the system and trigger the alarm, with a confidence level 
given by the Chernoff bound.

The $N$-pass IFM apparatus itself is based on a polarization interferometer that operates
in the basis of linear polarizations $|\rm{H}\rangle$ and $|\rm{V}\rangle$. The path of vertical polarization
constitutes a tripwire. The evolution of a photon's polarization state is described 
by successive multiplication of matrices $\hat{U}\left(\theta_N\right)$, $\hat{L}(\lambda)$, $\hat{O}(h)$ corresponding to 
polarization rotation by $\theta_N$ and loss, $\lambda$, of a photon in the detection arm:
\begin{equation}
\hat{U}\left(  \theta_{N}\right)  =%
\left[\matrix{
\cos\theta_{N} & -\sin\theta_{N}\cr
\sin\theta_{N} & \cos\theta_{N}\cr
}\right]\qquad
{\rm  and }\qquad\hat{L}\left(  \lambda\right)  =%
\left[\matrix{
1 & 0\cr
0 & \sqrt{1-\lambda}\cr
}\right] ,
\end{equation}
as well as the presence $\hat{O}(h_1)=\hat{L}(1)$ or absence $\hat{O}(h_0)=\hat{L}(0)$ of an object. If the 
input state of a photon is $|\psi_0\rangle$ then after a single pass it will be
$|\psi_1\rangle=\hat{O}(h)\hat{L}\hat{U}\left(\theta_N\right)
|\psi_0\rangle$. The probability to detect a photon with polarization X 
after $N$ passes is $P_{\rm{X}}=\langle\psi_N|\rm{X}\rangle\langle \rm{X}|\psi_N\rangle$, while the probability of total transmission
is $P_{\rm{tr}}=\langle\psi_N|\psi_N\rangle$, where $|\psi_N\rangle$ is obtained by repeating a single-pass evolution $N$ 
times.


\begin{figure}[t]
\begin{center}
\includegraphics{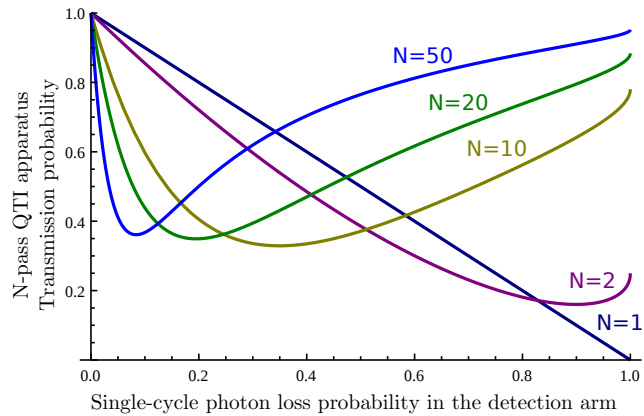}
\end{center}
\caption{\label{fig:Transmissions} The single-photon transmission probability
in a $N$-pass IQT apparatus $P_{\rm{tr}}$ for $N\theta_N=\pi/2$ as a function of single-cycle 
probability of photon loss in the detection arm. Loss in a $N$-pass IQT is optimized for this partial
Zeno effect to take place. The detection of an object is based on increase of transmission.
}
\end{figure}

In the IFM apparatus, a photon is initially horizontally polarized, 
$|\psi_0\rangle=|\rm{H}\rangle$. 
With each pass, polarization is rotated by an angle $\theta_N$, which increases a
photon's probability to be in the detection arm, where the photon interacts
with a beam splitter before being sent along the tripwire. 
We present the transmission probability $P_{\rm tr}$ as a function of a single-cycle 
probability of photon loss in the detection arm, $\lambda$, in the absence of an object 
(see figure~\ref{fig:Transmissions}). $P_{\rm tr}$ is given for a different number of passes
but with the same angle of evolution $N\theta_N=\pi/2$. A 100\% photon loss corresponds 
to the presence of an object in the detection arm. One can see, transmission in 
this case improves with the number of passes due to the quantum Zeno effect. The 
region of small $\lambda$ demonstrates how an artificial lossless case behaves since 
even a small amount leads to a significant drop in the transmission probability. 
Interestingly, the smallest transmission probability is for relatively high loss,
but it is not high enough for the quantum Zeno effect to become apparent. This 
partial Zeno effect corresponds to a special type of quantum state evolution in 
the presence of a probabilistic measurement. 

Our quantum interrogation technique is based on this special evolution. 
A controllable amount of loss $\lambda$ is introduced in the detection arm by 
means of a beam splitter with tunable reflectivity. This additional loss in the 
presence of an object reduces the probability of a photon striking the object 
during a trial, $P_{\rm{str}}=(1-\lambda)(1-\cos^{2N}\theta_N)$. 
Furthermore, we assume that reflectivity and phase shift
of the additional beam splitter (inside the interferometer) are 
constantly adjusted such that 
detection of a photon at the output is minimal---in order to
operate the device at the minimum of the curve
shown in figure~\ref{fig:Transmissions}. 
Such an adjustment is made in 
order to counteract changes in the environment as well as for the partial Zeno effect
to be maintained, which would obviously not be possible in the presence of 
an object. Thus hypotheses testing is based on two outcomes: a low
probability to detect a photon at the output in the absence of an object and 
100\% in its presence.

The Chernoff distance, in the case of
a hypotheses testing apparatus with only two outcomes, registered with 
probabilities $p_1(1)=p$ and $p_1(2)=\bar{p}$ or $p_0(1)=q$ and $p_0(2)=\bar{q}$, is
\begin{equation}
C_2(p,q)=\xi  \ln  \frac{\xi }{p}+\bar{\xi } \ln  \frac{\bar{\xi }}{\bar{p}},
\end{equation}
where $\xi =\ln \left(\bar{q}/\bar{p}\right)/\left(\ln \left(p/\bar{p}\right)+
\ln \left(\bar{q}/q\right)\right)$ and $\bar{x}=1-x$. Therefore, knowing $p$ and $q$ is 
sufficient for error estimation. The transmission probability could be 
calculated analytically only in the presence of the object, $p=\cos^{2N}\theta_N$. However,
in the absence of an object, the transmission probability, $q$, is 
experimentally available information, which is constantly provided by the IQT 
apparatus. 

There are two primary goals of the IQT apparatus: detection of an object with 
high certainty, $P_e\rightarrow 0$, while staying invisible, $\bar{P}_{\rm{vis}}(M)\approx 1$. 
Although satisfying both goals is, in principle, possible (see figure \ref{fig:result}), 
its success is limited by the number of passage $N$ performed in practice. 
Thus the following compromise between confidence level and invisibility is assumed. 
We would like $\bar{P}_{\rm{vis}}(M)>P_e$ (in fact $\bar{P}_{\rm{vis}}(M)>P_e^{\rm max}$), 
which means a higher
likelihood of not hitting the object with a photon than accepting the wrong 
hypothesis, while maintaining a confidence level above a blind guess: $1-P_e^{\rm max}>0.5$.

\section{Results}
In our apparatus, it is assumed that a tripwire becomes visible  after a single 
event of a photon striking an object. Therefore, the probability of a tripwire 
to stay invisible after $M$ trials is $\bar{P}_{\rm{vis}}(M)=\exp \left(-M C_{\rm{vis}}\right)$ as before where the 
visibility distance,  $C_{\rm{vis}}=-\ln (1-P_{\rm{str}})$, is defined in terms of the probability 
to strike an object, as described earlier. 

\begin{figure}[t]
\begin{center}
\includegraphics{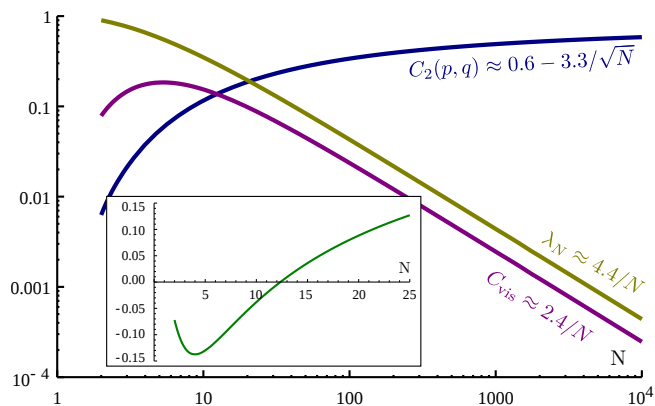}
\end{center}
\caption{\label{fig:result}
Chernoff $C_2(p,q)$ and visibility $C_{\rm{vis}}$ distances as a 
function of number of passes $N$ as
well as the amount of loss, $\lambda_N$ in the detection arm required for 
partial Zeno to take place. Inset is a difference between those distances. 
Invisible detection becomes possible when this difference becomes positive.
}
\end{figure}

We numerically simulated the performance of the IQT apparatus based on the state
evolution described above. 
For a given number of passes $N$ and $\theta_N$, 
we numerically found the optimal value of loss $\lambda_N$ that minimizes 
the single-trial transmission probability 
($P_{\rm{tr}}$ in the absence of an object). 
Then we used this value ($\lambda = \lambda_N$) to calculate the Chernoff and 
visibility distances $C_2(p,q)$ and $C_{\rm{vis}}$. 
Figure~\ref{fig:result} summarizes these 
results for a total angle of evolution $N\theta_N=\pi/2$ as a function of number of 
passes. This reveals that at least 13 passes are necessary for visibility 
distance to become smaller than Chernoff distance thus allowing for 
$\bar{P}_{\rm{vis}}(M)>2P_e(M)$.
While the IQT operating at $N = 100$ is experimentally feasible
with current technology,
the plots are extended 
over the range $N >100$ in order to demonstrate the asymptotic behavior.

\begin{figure}[b]
\begin{center}
\includegraphics{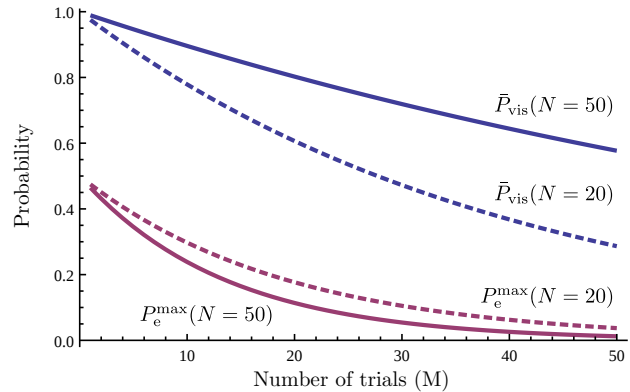}
\end{center}
\caption{\label{fig:result2}
Probability for the tripwire being invisible $\bar{P}_{\rm{vis}}$
and the maximum error bound $P_e^{\rm max}$ are given  
as functions of the number of trials $M$,
for given number of passes $N=20$ and $N=50$. 
As $N$ gets larger, $\bar{P}_{\rm{vis}}$ will stay closer to one
while $P_e^{\rm max}$ will go faster to zero.
}
\end{figure}

The Chernoff and visibility distances are directly translated to the
probability for the tripwire being invisible 
$\bar{P}_{\rm{vis}}(M)=\exp \left(-M C_{\rm{vis}}\right)$,
and the maximum error bound of probability making the wrong decision 
$P_e^{\rm max}(M)=\frac{1}{2} \exp \left(-M C_{2}(p,q)\right)$.
Figure~\ref{fig:result2} shows 
the dependence of $\bar{P}_{\rm{vis}}$ and $P_e^{\rm max}$ on 
the number of trials $M$, for given numbers of passes $N=20$ and $N=50$.
We note that, as the number of passes $N$ gets larger, 
$\bar{P}_{\rm{vis}}$ stays closer to one and
$P_e^{\rm max}$ goes faster to zero---allowing the ideal IQT.

\begin{table}[t]
\caption{\label{tab:results}
Ratio of the distances, visibility distance with a corresponding 
controllable loss for two cases of $N\theta_N$.}
\begin{indented}
\item[]
\begin{tabular}{ccccccc}
\br
& \multicolumn{3}{c}{$N\theta_N=\pi/2$}&\multicolumn{3}{c}{$N\theta_N=\pi/4$}\\
\mr
N & $\frac{C_{2}\left(  p,q\right)}{C_{\rm{vis}}\left(  N\right) }  $ & $C_{\rm{vis}}\left(  N\right)  $ & $\lambda
$ & $\frac{C_{2}\left(  p,q\right)}{C_{\rm{vis}}\left(  N\right) }  $ & $C_{\rm{vis}}\left(  N\right)  $ & $\lambda
$\\
\mr
5 & 0.29 & 0.184 & 0.575& 0.28 & 0.057 & 0.523\\
10 & 0.75 & 0.154 & 0.349& 0.79 & 0.042 & 0.314\\
11 & 0.85 & 0.147 & 0.324& 0.92 & 0.039 & 0.291\\
12 & 0.96 & 0.140 & 0.302& 1.00 & 0.038 & 0.271\\
13 & 1.07 & 0.133 & 0.282& 1.14 & 0.035 & 0.253\\
20 & 1.91 & 0.098 & 0.195& 2.08 & 0.025 & 0.174\\
50 & 6.16 & 0.045 & 0.084& 6.73 & 0.011 & 0.075\\
\br
\end{tabular}
\end{indented}
\end{table}

Table \ref{tab:results} presents numerical values of the visibility distance, 
the ratio of the distances, as well as the operational amount of loss in the 
detection arm, $\lambda_N$. It again shows that at least 13 passes are required before 
the statistical error starts going to zero faster than the probability of 
staying invisible. It also shows that a requirement of the total angle of 
rotation to be $N\theta_N=\pi/2$, which is a requirement for the standard $N$-pass IFM
apparatus, could be dropped. One can actually use $\theta_N$ as an additional parameter
for the optimization of IQT apparatus. In the case of $\pi/4$, the visibility 
distance is shortened by a factor of four. 
The shorter the distance the 
more trials are necessary, thus allowing for longer acquisition times (with larger $M$) 
and better averaging out of any additional errors acquired in a single trial. 
In addition, one can see that the 
Chernoff distance actually becomes greater relative to the visibility distance, 
which signifies that for the same probability of invisibility, statistical error
could be made smaller for the $\pi/4$ case than it was possible with a greater total
angle of rotation. Finally, the amount of controlled loss in the detection arm
is relatively high, which is comforting for practical realizations.

\section{Conclusion}
In conclusion, we have presented an IQT apparatus that is robust against both 
loss of photons and random phase accumulations in the detection arm due to
a built-in feedback. Interaction-free 
hypotheses testing in an IQT apparatus allows for stealth operation: 
detection of an intrusion while being virtually undetectable by an intruder.  
In addition, our apparatus does not require analyzing a photon's polarization 
state and does not rely on an exact $\pi/2$ rotation, thus allowing for the fine 
tuning of the performance. Therefore such an IQT apparatus holds great promise 
for practical applications related to security.

\ack
This work was supported by the 
Army Research Office, the Boeing corporation, the Department of Energy, the 
Foundational Questions Institute, the Intelligence Advance Research Projects 
Activity, and the Northrop-Grumman corporation.


\end{document}